\documentclass[sigconf, printfolios=true]{acmart}
\pdfoutput=1

\definecolor{rwthblue}   {RGB}{  0,  84, 159}
\definecolor{rwthblue75}{RGB}{ 64, 127, 183}
\definecolor{rwthblue50}{RGB}{142, 186, 229}
\definecolor{rwthblue25}{RGB}{199, 221, 242}
\definecolor{rwthblue10}{RGB}{232, 241, 250}

\definecolor{rwthblack}  {RGB}{  0,   0,   0}
\definecolor{rwthblack75}{RGB}{100, 101, 103}
\definecolor{rwthblack50}{RGB}{156, 158, 159}
\definecolor{rwthblack25}{RGB}{207, 209, 210}
\definecolor{rwthblack10}{RGB}{236, 237, 237}

\definecolor{rwthdarkgray}{named}{rwthblack75}
\definecolor{rwthgray}{named}{rwthblack50}
\definecolor{rwthlightgray}{named}{rwthblack25}
\definecolor{rwthverylightgray}{named}{rwthblack10}

\definecolor{rwthmagenta}  {RGB}{227,   0, 102}
\definecolor{rwthmagenta75}{RGB}{233,  96, 136}
\definecolor{rwthmagenta50}{RGB}{241, 158, 177}
\definecolor{rwthmagenta25}{RGB}{249, 210, 218}
\definecolor{rwthmagenta10}{RGB}{253, 238, 240}

\definecolor{rwthyellow}  {RGB}{255, 237,   0}
\definecolor{rwthyellow75}{RGB}{255, 240,  85}
\definecolor{rwthyellow50}{RGB}{255, 245, 155}
\definecolor{rwthyellow25}{RGB}{255, 250, 209}
\definecolor{rwthyellow10}{RGB}{255, 253, 238}

\definecolor{rwthpetrol}   {RGB}{  0,  97, 101}
\definecolor{rwthpetrol75}{RGB}{ 45, 127, 131}
\definecolor{rwthpetrol50}{RGB}{125, 164, 167}
\definecolor{rwthpetrol25}{RGB}{191, 208, 209}
\definecolor{rwthpetrol10}{RGB}{230, 236, 236}

\definecolor{rwthturquoie}   {RGB}{  0, 152, 161}
\definecolor{rwthturquoise75}{RGB}{  0, 177, 183}
\definecolor{rwthturquoise50}{RGB}{137, 204, 207}
\definecolor{rwthturquoise25}{RGB}{202, 231, 231}
\definecolor{rwthturquoise10}{RGB}{235, 246, 246}

\definecolor{rwthgreen}  {RGB}{ 87, 171,  39}
\definecolor{rwthgreen75}{RGB}{141, 192,  96}
\definecolor{rwthgreen50}{RGB}{184, 214, 152}
\definecolor{rwthgreen25}{RGB}{221, 235, 206}
\definecolor{rwthgreen10}{RGB}{242, 247, 236}

\definecolor{rwthlightgreen}   {RGB}{189, 205,   0}
\definecolor{rwthlightgreen75}{RGB}{208, 217,  92}
\definecolor{rwthlightgreen50}{RGB}{224, 230, 154}
\definecolor{rwthlightgreen25}{RGB}{240, 243, 208}
\definecolor{rwthlightgreen10}{RGB}{249, 250, 237}

\definecolor{rwthorange}  {RGB}{246, 168,   0}
\definecolor{rwthorange75}{RGB}{250, 190,  80}
\definecolor{rwthorange50}{RGB}{253, 212, 143}
\definecolor{rwthorange25}{RGB}{254, 234, 201}
\definecolor{rwthorange10}{RGB}{255, 247, 234}

\definecolor{rwthred}  {RGB}{204,   7,  30}
\definecolor{rwthred75}{RGB}{216,  92,  65}
\definecolor{rwthred50}{RGB}{230, 150, 121}
\definecolor{rwthred25}{RGB}{243, 205, 187}
\definecolor{rwthred10}{RGB}{250, 235, 227}

\definecolor{rwthbordeaured}   {RGB}{161,  16,  53}
\definecolor{rwthbordeauxred75}{RGB}{182,  82,  86}
\definecolor{rwthbordeauxred50}{RGB}{205, 139, 135}
\definecolor{rwthbordeauxred25}{RGB}{229, 197, 192}
\definecolor{rwthbordeauxred10}{RGB}{245, 232, 229}

\definecolor{rwthviolet}  {RGB}{ 97,  33,  88}
\definecolor{rwthviolet75}{RGB}{131,  78, 117}
\definecolor{rwthviolet50}{RGB}{168, 133, 158}
\definecolor{rwthviolet25}{RGB}{210, 192, 205}
\definecolor{rwthviolet10}{RGB}{237, 229, 234}

\definecolor{rwthpurple}  {RGB}{122, 111, 172}
\definecolor{rwthpurple75}{RGB}{155, 145, 193}
\definecolor{rwthpurple50}{RGB}{188, 181, 215}
\definecolor{rwthpurple25}{RGB}{222, 218, 235}
\definecolor{rwthpurple10}{RGB}{242, 240, 247}

\newcommand{\sym}[1]{\text{Symmetric}\left( #1 \right)}

\newcommand{\lotri}[1]{\text{lowerTriangular}\left( #1 \right)}
\newcommand{\uptri}[1]{\text{upperTriangular}\left( #1 \right)}

\newcommand{\diag}[1]{\text{Diagonal}\left( #1 \right)}

\usepackage{tikz}
\usetikzlibrary{positioning}
\usetikzlibrary{shapes.geometric}
\usetikzlibrary{plotmarks}
\tikzset{tbox/.style={fill=rwthblack10, rounded corners}}
\tikzset{active/.style={fill=rwthblue50, rounded corners}}
\tikzset{barrow/.style={line width=0.2mm}}

\usepackage{pgfplots}
\usepgfplotslibrary{statistics}
\usepackage{pgfplotstable}
\pgfplotsset{compat=1.14}
\pgfplotsset{x tick label style={/pgf/number format/1000 sep=}}

\usepackage{listings}
\lstdefinestyle{mystyle}{
	language=Python,
	columns=[c]fullflexible,
	mathescape=true,
	tabsize=2,
	breaklines=true,
	lineskip=1pt,
	morekeywords={yield},
	escapeinside={(*}{*)}
}
\newcommand{\su}{{\footnotesize$\times$}}

\def\BibTeX{{\rm B\kern-.05em{\sc i\kern-.025em b}\kern-.08emT\kern-.1667em\lower.7ex\hbox{E}\kern-.125emX}}

\copyrightyear{2020} 
\acmYear{2020} 
\setcopyright{acmcopyright}\acmConference[PASC '20]{Proceedings of the Platform for Advanced Scientific Computing Conference}{June 29-July 1, 2020}{Geneva, Switzerland}
\acmBooktitle{Proceedings of the Platform for Advanced Scientific Computing Conference (PASC '20), June 29-July 1, 2020, Geneva, Switzerland}
\acmPrice{15.00}
\acmDOI{10.1145/3394277.3401836}
\acmISBN{978-1-4503-7993-9/20/06}

\begin{document}

%
\title{Automatic Generation of Efficient Linear Algebra Programs}

%
\author{Henrik Barthels}
\affiliation{%
  \institution{AICES, RWTH Aachen University}
  \streetaddress{Schinkelstr. 2}
  \city{Aachen}
  \postcode{52062}
  \country{Germany}
}
\email{barthels@aices.rwth-aachen.de}

\author{Christos Psarras}
\affiliation{%
  \institution{AICES, RWTH Aachen University}
  \streetaddress{Schinkelstr. 2}
  \city{Aachen}
  \postcode{52062}
  \country{Germany}
}
\email{psarras@aices.rwth-aachen.de}

\author{Paolo Bientinesi}
\affiliation{%
  \institution{Ume\aa{} Universitet}
  \city{Ume\aa{}}
  \country{Sweden}
}
\email{pauldj@cs.umu.se}

%

%
\begin{abstract}
The level of abstraction at which application experts reason about linear algebra computations and the level of abstraction used by developers of high-performance numerical linear algebra libraries do not match. The former is conveniently captured by high-level languages and libraries such as Matlab and Eigen, while the latter expresses the kernels included in the BLAS and LAPACK libraries.  Unfortunately, the translation from a high-level computation to an efficient sequence of kernels is a task, far from trivial, that requires extensive knowledge of both linear algebra and high-performance computing. Internally, almost all high-level languages and libraries use efficient kernels; however, the translation algorithms are too simplistic and thus lead to a suboptimal use of said kernels, with significant performance losses.  In order to both achieve the productivity that comes with high-level languages, and make use of the efficiency of low level kernels, we are developing Linnea, a code generator for linear algebra problems. As input, Linnea takes a high-level description of a linear algebra problem and produces as output an efficient sequence of calls to high-performance kernels. In 25 application problems, the code generated by Linnea always outperforms Matlab, Julia, Eigen and Armadillo, with speedups up to and exceeding 10\su{}.
\end{abstract}

%
%


%

%

%
\maketitle

\section{Introduction}

A common high-performance computing workflow to accelerate the execution of target application problems
consists in first identifying a set of computational building blocks, and then engaging in extensive algorithmic and code optimization.
Although this process leads to sophisticated and highly-tuned code, the performance gains in the computational building blocks do not necessarily carry over to the high-level application problems domain experts solve in their day-to-day work.

In the domain of linear algebra, significant effort is put into optimizing BLAS and LAPACK implementations for all the
different architectures and generations, and  for operations such as matrix-matrix multiplication, nearly optimal efficiency rates are attained. However, we observe a decrease in the number of users that actually go through the tedious, error-prone and time consuming process of using directly said libraries by writing their code in C or Fortran; instead, languages and libraries such as Matlab, Julia, Eigen and Armadillo, which offer a higher level of abstraction, are becoming more and more popular.
These languages and libraries allow users to input a linear algebra problem as an expression which closely resembles the
mathematical description; this expression is then internally mapped to lower level building blocks such as BLAS and LAPACK.
Unfortunately, our experience suggests that this translation frequently results in suboptimal code.

The following examples illustrate some of the challenges that arise in the mapping from high-level expression to
low-level kernels. A straightforward translation of the assignment $y_k := H^\dag y + ( I_n - H^\dag H ) x_k$, which
appears in an image restoration application \cite{Tirer:2017uv}, will result in code containing one $\mathcal{O}(n^3)$
matrix-matrix multiplication to compute $H^\dag H$. By contrast, by means of distributivity, this assignment can be
rewritten as $y_k := H^\dag (y - H x_k) + x_k$, and computed with only $\mathcal{O}(n^2)$ matrix-vector
multiplications. The computation of the expression
$$\frac{k}{k-1}B_{k-1} (I_n - A^T W_k ((k-1)I_l + W_k^T A B_{k-1} A^T W_k)^{-1} W_k^T A B_{k-1} )\text{,}$$
which is part of a stochastic method for the solution of least squares problems \cite{Chung:2017ws}, can be sped up by
identifying that the subexpression $W_k^T A$ or its transpose $(A^T W_k)^T$ appear four times. Often times,
application experts possess domain knowledge that leads to better implementations: In $x := (A^T A + \alpha^2 I)^{-1} A^T b$ \cite{Golub:2006hl}, since $\alpha > 0$, it can be inferred that $A^T A + \alpha^2 I$ is always symmetric positive definite. As a result, the linear system can always be solved by using the relatively inexpensive Cholesky factorization. In most languages and libraries, such additional knowledge about properties of the operands either can not be specified, or it is not automatically exploited.

In this paper, we discuss Linnea, a prototype of a code generator 
that automates the translation of the mathematical description of a linear algebra problem to an efficient sequence of calls to BLAS and LAPACK kernels.\footnote{
Linnea is available at \url{https://github.com/HPAC/linnea}.}
Linnea is written in Python and targets mid-to-large scale linear algebra expressions, where problems are typically compute bound. One of the advantages of Linnea is that all optimizations are performed symbolically, using rewrite rules and term replacement, so the generated programs are correct by construction.

Linna currently supports real-valued computations, and parallelism via multi-threaded kernels. Its input grammar is shown in Fig.~\ref{fig:grammar}. Matrices can be annotated with the following properties: upper or lower triangular, diagonal, symmetric, symmetric positive definite (SPD), symmetric positive semi definite, and orthogonal.
As building blocks, Linnea uses BLAS and LAPACK kernels, as well as a small number of code snippets for simple
operations not supported by those libraries. As output, we decided to generate Julia code because it offers a good
tradeoff between simplicity and performance: Low-level wrappers expose the full functionality of BLAS and LAPACK, while
additional routines can be implemented easily without compromising performance \cite{Bezanson:2018ip}.

Experiments indicate that the code generated by Linnea outperforms all standard linear algebra languages and libraries. At the same time, the code generation time is mostly in the order of a few minutes, that is, significantly faster than human experts.

\begin{figure}[]
\begin{minipage}{\columnwidth}
\begin{align*}
	\text{assignments} &\rightarrow \text{assignment}^{+}\\
	\text{assignment} &\rightarrow \text{symbol} := \text{expr}\\
	\text{expr} &\rightarrow \text{symbol} \mid \text{expr} + \text{expr} \mid \text{expr} \cdot \text{expr} \mid\\
		&\qquad  - \text{expr} \mid \text{expr}^{-1} \mid \text{expr}^{T} \mid (\text{expr}) \\
	\text{symbol} &\rightarrow \textit{matrix} \mid \textit{vector} \mid \textit{scalar}
\end{align*}
\end{minipage}
    \caption{The grammar for the input expressions of Linnea. In addition, operands can be annotated with properties.}
    \label{fig:grammar}
\end{figure}

\paragraph{Organization of the paper}
Sec.~\ref{sec:relatedwork} surveys the related work. The basic ideas behind the code generation in Linnea are introduced in Sec.~\ref{sec:synthesis}. Details of the implementation are discussed in Sec.~\ref{sec:implementation}. Experimental results are presented in Sec.~\ref{sec:experiments}. Sec.~\ref{sec:conclusion} concludes the paper.

\section{Related Work}
\label{sec:relatedwork}

\subsection{Code Generation}

The problem of translating the intermediate representation of a program in the form of an expression tree to machine instructions is closely related to the problem discussed in this paper. However, existing approaches using pattern matching and dynamic programming \cite{Aho:1976ga, aho1989}, as well as bottom-up rewrite systems (BURS) \cite{pelegri1988} solely focus on expressions containing basic operations directly supported by machine instructions. The two main objectives of code generation are to minimize the number of instructions and to use registers optimally. While there are approaches that generate optimal code for arithmetic expressions, considering associativity and commutativity \cite{sethi1970}, more complex properties of the underlying algebraic domain, for example distributivity, are usually not considered.

Instead of applying optimization passes sequentially, Equality Saturation (EQ) \cite{Tate:2009kz} is a compilation technique that constructs many equivalent programs simultaneously, stored in a single intermediate representation. Domain specific knowledge can be provided in the form of axioms. Equality Saturation is more general in its scope than Linnea, as it allows for control flow. So far, EQ has only been implemented for Java, and it is not clear how well it would scale with the large number of axioms required to encode the optimizations that Linnea carries out.

If one wishes to integrate Linnea into an existing language, Leightweight Modular Staging \cite{Rompf:2011he}, a technique to implement an optimizing compiler framework as a library within a host language, could be used.

\subsection{Tools and Languages for Linear Algebra}

Presently, a range of tools are available for the computation of linear algebra expressions. At one end of the spectrum
there are the aforementioned high-level programming languages such as Matlab, Julia, R and Mathematica. In those languages, working code can be written within minutes, with little or no knowledge of numerical linear algebra. However, the resulting code (possibly numerically unstable\footnote{Some systems compute the condition number for certain operations and give a warning if results may be inaccurate.}) usually achieves suboptimal performance. One of the reasons is that, with the exception of a small number of functions, these tools do not make it possible to express properties. A few Matlab functions exploit properties by inspecting matrix entries, a step which could be avoided with a more general method to annotate operands with properties. Furthermore, if the inverse operator is used, an explicit inversion takes place, even if the faster and numerically more stable solution would be to solve a linear system \cite[Sec. 13.1]{higham1996}; it is up to the user to rewrite the inverse in terms of operators that solve linear systems, for example ``$\slash$'' or ``$\backslash$'' in Matlab \cite{matlabdoc:short}.

At the other end of the spectrum there are C/Fortran libraries such as BLAS \cite{dongarra1990:short} and LAPACK
\cite{anderson1999:short}, which offer highly optimized kernels for basic linear algebra operations. However, the
translation of a target problem into an efficient sequence of kernel invocations is a lengthy and error-prone process that requires deep understanding of both numerical linear algebra and high-performance computing.

In between, there are expression template libraries such as Eigen \cite{eigenweb}, Blaze \cite{Iglberger:2012hb}, and
Armadillo \cite{sanderson2010}, which provide a domain-specific language integrated within C++. They offer a compromise
between ease of use and performance. The main idea is to improve performance by eliminating temporary operands. While
both high-level languages and libraries increase the accessibility, they almost entirely ignore domain knowledge, and
because of this, they frequently deliver slow code.

The Transfor program \cite{gomez1998} is likely the first translator of linear algebra problems (written in Maple) into BLAS kernels; since the inverse operator was not supported, the system was only applicable to the simplest expressions.  More recently, several other solutions to different variants of this problem have been developed: The Formal Linear Algebra Methods Environment (FLAME) \cite{Gunnels:2001gi,Bientinesi:2005hu} is a methodology for the derivation of algorithmic variants for basic linear algebra operations such as factorizations and the solution of linear systems. Cl1ck \cite{FabregatTraver:2011km,FabregatTraver:2011gu} is an automated implementation of the FLAME methodology.  The goal of BTO BLAS is to generate C code for bandwidth bound operations, such as fused matrix-vector operations \cite{Siek:2008ij}.  In contrast to the linear algebra compiler CLAK \cite{fabregat-traver2013a:short}, which inspired the presented code generation approach, we make use of the algebraic nature of the domain to remove redundancy during the derivation.  DxTer uses domain knowledge to optimize programs represented as dataflow graphs \cite{marker2012, marker2015}.  LGen targets basic linear algebra operations for small operand sizes, a regime in which BLAS and LAPACK do not perform very well, by directly generating vectorized C code \cite{spampinato2016:short}.  SLinGen \cite{Spampinato:2018tz} combines Cl1ck and LGen to generate code for more complex small-scale problems, but still requires that the input is described as a sequence of basic operations.
Similar approaches for code generation exist for related domains such as tensor contractions (TCE \cite{Baumgartner:2005dq}) and linear transforms (Spiral \cite{Franchetti:eq}, FFTW \cite{Frigo:2005cp}).

Our aim is to combine the advantages of existing approaches: The simplicity, and thus, productivity, of a high-level language, paired with performance that comes close to what is achieved manually by human experts.

\section{Algebraic Code Generation}
\label{sec:synthesis}

The core idea behind Linnea is to rewrite the input problem while successively identifying parts of the problem that are computable by a sequence of one or more available kernels. In order to efficiently store the large number of generated sequences, which all compute the input, but differ with respect to a given cost function, we use a graph: Nodes represent the input at different stages of the computation, and edges are annotated with the sequences to transition from one stage to another.

Algebraic code generation starts with a single root node containing a symbolic expression that represents the input problem. This expression is rewritten in different ways, for example making use of distributivity. The different representations of a given expression are not stored explicitly; instead a node only contains one variant, and this variant is rewritten when necessary. On each variant of the expression, pattern matching is used to search for subexpressions that can be computed with the available kernels. Whenever a match is found, a new successor of the root node is constructed. The edge from the root to the new child node is annotated with the kernel. This child contains the expression that is left after using the kernel to compute part of the original input.

The two steps of rewriting expressions and pattern matching are then repeated on the new nodes, until nodes are found with nothing left to compute. In practice, this process corresponds to the construction and traversal of a graph. Linnea currently uses a breadth-first implementation, but it would also be possible to use a depth-first approach.

Upon termination, the concatenation of all kernels along a path in the graph from the root to a leaf is a program that computes the input problem. Given a function that assigns a cost to each kernel, the optimal program is found by searching for the shortest path in the graph from the root node to a leaf.

\subsection{The Algorithm}

\begin{figure}
\hrule
\begin{lstlisting}
$V = V_\text{active} = \{v_\text{input}\}$
while $V_\text{active} \neq \varnothing$ and $|V_\text{terminal}| < \text{threshold}$:
	$V_\text{next} := \varnothing$
	for $v$ in $V_\text{active}$:
		for $v_r$ in representations($v$):
			$V_\text{new} := \text{gen\_successors}(v_r)$
			$V_\text{next} := V_\text{next} \cup V_\text{new}$
			$V := V \cup V_\text{new}$
	$V_\text{active} := V_\text{next}$
	# merge nodes (see next Section)
# find optimal path on $G = (V, E)$ according to cost function
\end{lstlisting}
\hrule
\caption{Pseudocode for the code generation algorithm.}
\label{fig:pseudocode}
\end{figure}

In practice, algebraic code generation can be implemented by maintaining a set of so called \emph{active nodes} $V_\text{active}$, as illustrated in Fig.~\ref{fig:pseudocode}. This set can be seen as the front or the leaves of $G$. In every iteration of the while loop, new successors are generated from the current set of active nodes, which will become the set of active nodes in the next iteration. At the same time, the current set of active nodes will become inactive. The two steps of rewriting and matching are implemented in two functions: The `gen\_successors' function uses pattern matching and generates a set of new successors for a single node. The `representations' function rewrites terms in different ways. As an example, for $A(B+C)$, it returns the original term $A(B+C)$, as well as $AB+AC$. We omitted the construction of edges from the pseudocode; it is assumed that this is done by `gen\_successors'.

The loop terminates either when there are no more active nodes ($V_\text{active} = \varnothing$), or optionally when a certain number of terminal nodes is found ($|V_\text{terminal}| \ge \text{threshold}$).

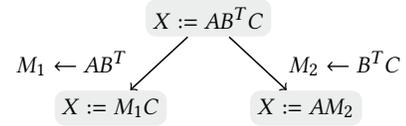
\begin{figure}
\begin{tikzpicture}[node distance=1cm]

\node (n1) [tbox]  {$X := A B^T C$};

\node (n2a) [tbox, below left=of n1, xshift=1cm]  {$X := M_1 C$};

\node (n2b) [tbox, below right=of n1, xshift=-1cm]  {$X := A M_2$};

\path[->] (n1) edge [barrow] node [left, xshift=-0.3cm] {$M_1 \leftarrow A B^T$} (n2a);

\path[->] (n1) edge [barrow] node [right, xshift=0.3cm] {$M_2 \leftarrow B^T C$} (n2b);

\end{tikzpicture}
\caption{Derivation graph for $X := A B^T C$ after one iteration.}
\label{fig:dgbefore}
\end{figure}

\begin{figure}
\begin{tikzpicture}[node distance=1cm]

\node (n1) [tbox]  {$X := A B^T C$};

\node (n2a) [tbox, below left=of n1, xshift=1cm]  {$X := M_1 C$};

\node (n2b) [tbox, below right=of n1, xshift=-1cm]  {$X := A M_2$};

\node (n3a) [tbox, below=of n2a, yshift=0.2cm]  {$X := M_3$};

\node (n3b) [tbox, below=of n2b, yshift=0.2cm]  {$X := M_4$};

\path[->] (n1) edge [barrow] node [left, xshift=-0.3cm] {$M_1 \leftarrow A B^T$} (n2a);

\path[->] (n1) edge [barrow] node [right, xshift=0.3cm] {$M_2 \leftarrow B^T C$} (n2b);

\path[->] (n2a) edge [barrow] node [left, xshift=-0.0cm] {$M_3 \leftarrow M_1 C$} (n3a);

\path[->] (n2b) edge [barrow]  node [right, xshift=-0.0cm] {$M_4 \leftarrow A M_2$} (n3b);

\end{tikzpicture}
\caption{Derivation graph after two iterations.}
\label{fig:dgafter}
\end{figure}
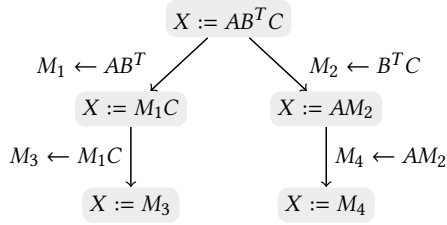

\begin{example}
  Let us assume the input is the expression $X := A B^T C$, and there are available kernels that compute $X Y$, $X^T Y$ and $X Y^T$. Initially, the derivation graph $G$ only contains a single root node with $X := A B^T C$. In the first iteration, matches are found for $X^T Y$ and $X Y^T$, so two new nodes are constructed. The resulting graph is shown in Fig.~\ref{fig:dgbefore}. $M_1$ and $M_2$ are intermediate operands that symbolically represent the result of applying $X Y^T$ and $X^T Y$, respectively. After the second iteration, the graph shown in Fig.~\ref{fig:dgafter} is obtained. Since both leaves only contain assignments with a single operand on the right-hand side, there is nothing left to compute and both nodes are terminal nodes. The two paths in the final graph represent two programs that compute $X := A B^T C$.\qed
\end{example}

\subsubsection{Derivation Strategies}

The basic idea of algebraic code generation is very general; it does not prescribe how exactly to implement `gen\_successors'. This function can be implemented as simple or complex as one wishes: It may simply use brute force pattern matching to apply kernels, or it can use specialized algorithms to constructively generate a good sequence of kernel invocations for subexpressions with specific structure. It is also possible to use multiple simple `gen\_successors' functions, instead of a single, more complicated one. In our implementation, we observed that a fine granularity made it easy to combine and reuse `gen\_successors' functions for different derivation strategies (see Sec.~\ref{sec:constructive}).

\subsection{Redundancy in the Derivation Graph}
\label{sec:redundancy}

\begin{figure}
\begin{tikzpicture}[node distance=1cm]

\node (n1) [tbox]  {$X := A(B+C+DE)$};

\node (n2a) [tbox, below left=of n1, xshift=1.3cm, yshift=-0.7cm]  {$X := M_3 + ADE$};

\node (n2b) [tbox, below right=of n1, xshift=-1.3cm, yshift=-0.7cm]  {$X := M_5 + ADE$};

\path[->] (n1) edge [barrow] node [left, xshift=-0.6cm] {
	$\begin{aligned}
		M_1 &\leftarrow AB \\
		M_2 &\leftarrow AC \\
		M_3 &\leftarrow M_1 + M_2
	\end{aligned}$
	} (n2a);

\path[->] (n1) edge [barrow] node [right, xshift=0.4cm] {
	$\begin{aligned}
		M_4 &\leftarrow B + C \\
		M_5 &\leftarrow A M_4
	\end{aligned}$
	} (n2b);

\end{tikzpicture}
\caption{Derivation graph with redundancy.}
\label{fig:dgredundancy}
\end{figure}
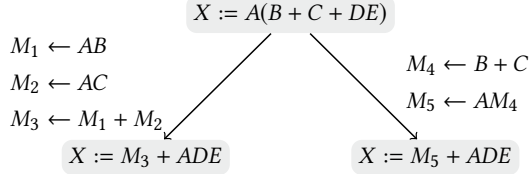

With large input expressions, it frequently happens that there is a lot of redundancy in the derivation graph. As an example, to compute the subexpressions $A(B+C)$ of $A(B+C+DE)$, the two different programs shown in Fig.~\ref{fig:dgredundancy} were constructed. As the generation process unfolds, both leaf nodes will be expanded, deriving the same programs for $ADE$ twice.

This phenomenon can be alleviated by taking advantage of the algebraic nature of the domain. In Fig.~\ref{fig:dgredundancy}, it is clear that $M_3$ and $M_5$ represent the same quantity because $AB+AC = A(B+C)$.\footnote{Ignoring differences due to floating-point arithmetic.} Thus, it is possible to merge the two branches and only do the derivation for $ADE$ once.

Our approach to detecting equivalent nodes and merging branch\-es in the derivation graph consists of two parts: First, we define a normal form for expressions, that is, a unique representation for algebraically equivalent terms. Second, we make sure that irrespective of how a subexpression was computed, its result is always represented by the same, unique intermediate operand. In case of the graph in Fig.~\ref{fig:dgredundancy}, this would mean that the same intermediate for $AB+AC = A(B+C)$ is used in both leaves. When rewritten to their normal form, the equivalence of two expressions can simply be checked by a syntactic comparison.

\subsubsection{Normal Form for Expressions}

As a normal form for linear algebra expressions, both a \emph{sum of products} (e.g., $AB + AC$) and a \emph{product of sums} (e.g., $A(B+C)$) can be used; we opted for the sum of products.\footnote{The reason is that it is not obvious how to make the product of sums form unique. As an example, the expression $AC + AD + BC$ can be written both as $A(C+D) + BC$ and $(A+B)C+AD$.} Terms in sums are sorted according to an arbitrary total ordering on terms. The transposition and inversion operators are pushed down as far as possible: As examples, the normal form of $(AB)^{-1}$ and $(A+B)^T$ is $B^{-1} A^{-1}$ and $A^T+B^T$, respectively. Since expressions are converted between different representations during the derivation, the normal form does not influence the quality of the generated code.

Deciding whether or not two different representations represent the same
element of an algebra is known as the \emph{word problem}, which in many cases is undecidable~\cite[pp. 59--60]{Baader:1999uu}. At least for some cases, this problem can be solved by a confluent and terminating term rewriting system, which can be obtained with Knuth-Bendix completion, or some of its extensions (for an overview, consider \cite{Dick:1991jh}). In practice, merging branches still works if some terms cannot correctly be identified as equivalent. This simply has the effect that some opportunities for merging will not be identified, so the optimization is less effective.

\subsubsection{Unique Intermediate Operands}

To ensure that the same intermediate operand is used for equivalent expressions, we make use of the normal form of expressions. The idea is to maintain a table of intermediate operands and the expressions they represent in the normal form. Whenever a kernel is used to compute part of an expression, we reconstruct the full expression that is computed by recursively replacing all intermediate operands. The resulting expression is then transformed to its normal form, and it is checked if there already is an intermediate operand for this expression in the table of intermediate operands.

\begin{example}

Let us assume we have the input $X:= A(B+C+D)$. Initially, the table of intermediate operands, which is shown in Tab.~\ref{tab:intermediates}, is empty. The first partial program is found by rewriting this assignment as $X:= AB+AC+AD$ and computing
$$T_1 \leftarrow AB \qquad T_2 \leftarrow AC \qquad T_3 \leftarrow T_1 + T_2\text{,}$$
%
resulting in $X := T_4 + AD$. For the first two kernels, we simply add the intermediates $T_1$ and $T_2$, and the
corresponding expressions $AB$ and $AC$ to the table. For $T_1 + T_2$, we first use the table to replace the
intermediate operands $T_1$ and $T_2$ with the expressions they represent, resulting in $AB+AC$. As this expression is
already in normal form, we can simply check if there already is an entry for it in the table. Since at this point, there
is no entry for $AB+AC$ yet, $AB+AC$ is added to the table, and a new operand $T_3$ is created.

Alternatively, the same part of $X:= A(B+C+D)$ can be computed as
$$T_4 \leftarrow B+C \qquad T_3 \leftarrow A T_4\text{.}$$
%
For the kernel invocation $A T_4$, the intermediate operand is created by replacing $T_4$ by $B+C$, and then converting the resulting expression $A(B+C)$ to normal form, which in this case is $AB+AC$. Then, from the table, $T_3$ is retrieved. Tab.~\ref{tab:intermediates} shows the state of the table after deriving those two programs.\qed
\begin{table}
\begin{tabular}{c c}
	\toprule
	intermediate & expression \\
	\midrule
	$T_1$ & $AB$ \\
	$T_2$ & $AC$ \\
	$T_3$ & $AB+AC$ \\
	$T_4$ & $B+C$ \\
	\bottomrule
\end{tabular}
\caption{The table of intermediate operands after deriving two programs computing the subexpression $A(B+C)$ in $X:= A(B+C+D)$.}
\label{tab:intermediates}
\end{table}
\end{example}

\subsubsection{Merging}

When merging branches, we implicitly assume that nodes do not have any state information such as the state of the registers or memory. This is a simplification that does not hold true in reality. However, without this assumption, it would not be possible to merge branches in the derivation graph. This optimization can drastically reduce the size of the derivation graph without reducing the size of the search space, allowing to derive programs for larger input expressions.

\section{Implementation}
\label{sec:implementation}

\subsection{Symbolic Expressions and Pattern Matching}

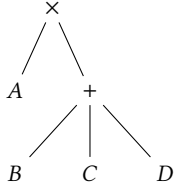
\begin{figure}
\begin{tikzpicture}[level distance=1.1cm, sibling distance=1cm]
	\node {$\times$}
		child {node {$A$}}
		child {node {$+$}
			child {node {$B$}}
			child {node {$C$}}
			child {node {$D$}}
		}
	;
\end{tikzpicture}
\caption{Expression tree for $A(B+C+D)$.}
\label{fig:exprtree}
\end{figure}

In Linnea, the input problem is represented as a symbolic expression, i.e., a tree-like algebraic data structure constructed from function symbols and constants, which represent operands. As an example, the expression $A(B+C+D)$ is represented by the tree shown in Fig.~\ref{fig:exprtree}. Instead of using nested binary operations, and thus nested binary expression trees, associative operations such as multiplication and addition are represented as n-ary operations.

Each available kernel is represented by a pattern, that is, a symbolic expressions with variables. As an example, the \texttt{gemm} kernel for matrix-matrix multiplication is described by the pattern $X Y + Z$, where $X$, $Y$ and $Z$ are variables that match a single matrix. To identify where kernels can be applied, we use associative-commutative pattern matching. A search for the pattern $X Y + Z$ in an expression then yields all subexpressions that can be computed with the \texttt{gemm} kernel.

Associative-commutative pattern matching allows to specify for each operator whether it is associative and/or
commutative. The pattern matching algorithms automatically takes those properties into account. As an example, with the
pattern $X^T +Y$ and the expression $A + B^T + C$, two matches are found: $B^T + A$ and $B^T + C$; since addition is
commutative, these two matches are found irrespective of how the terms in $A + B^T + C$ ordered.

To make use of specialized kernels that exploit the properties of matrices, we use patterns with constraints on the variables. A pattern only matches if the constraints for all operands are satisfied.

For pattern matching, we use the Python module MatchPy~\cite{krebber2017pyhpc, krebber2017scipy, krebber2018joss}, as it offers efficient many-to-one algorithms for this type of pattern matching~\cite{Krebber:2017tp}. For many-to-one matching, data structures similar to decision trees allow to make use of similarities between patterns to speed up matching.

\subsection{Matrix Properties}
\label{sec:properties}

Matrix properties are important to select the most fitting kernel to compute a given term. As an example, consider the product $A L$ where $A$ and $L$ are a full and a lower triangular matrix, respectively. This expression can be computed both with the \texttt{gemm} and the \texttt{trmm} kernels: The first one implements a general matrix-matrix product, while the second one requires one of the matrices to be triangular. Being more general, \texttt{gemm} performs twice as many floating-point operations as \texttt{trmm}~\cite[p. 336]{higham2008}, which should therefore be used instead whenever possible.

Linnea's input format makes it possible to annotate matrices with properties. However, not only is it important to know the properties of the input matrices, it is at least equally important to know the properties of intermediate operands, as the computation unfolds.
For this, we encoded linear algebra knowledge into a set of inference rules such as
\begin{align*}
	\lotri{A} &\rightarrow \uptri{A^T}\\
	\diag{A} \land \diag{B} &\rightarrow \diag{AB}\\
	A = A^T &\rightarrow \sym{A}\text{,}
\end{align*}
where $A$ and $B$ are arbitrary matrix expressions. As a trivial example, irrespective of how it is computed, the product of two lower triangular matrices yields another lower triangular matrix.

\subsection{Factorizations}

In contrast to other languages and libraries, in the input Linnea does not distinguish between the explicit matrix inversion and the solution of a linear system. Whenever possible, a linear system is solved; matrices are explicitly inverted only if this is unavoidable, for example in expressions such as $A^{-1} + B$. Even though LAPACK offers kernels that encapsulate a factorization and solve a linear system (e.g., \texttt{gesv}), Linnea ignores those kernels and directly applies factorizations. This is because the explicit factorization might enable other optimizations which are not possible when using a ``black box'' kernel such as \texttt{gesv}. As an example, consider the generalized least squares problem $b := (X^T M^{-1} X)^{-1} X^T M^{-1} y$, where $M$ is symmetric positive definite. This problem can be computed efficiently by applying the Cholesky factorization to $M$, resulting in $b := (X^T L^{-1} L^{-T} X)^{-1} X^T L^{-1} L^{-T} y$. In this expression, the subexpression $X^T L^{-1}$ or its transpose $L^{-T} X$ appear three times and only need to be computed once. If either $X^T M^{-1}$ or $M^{-1} X$ were computed with a single kernel, this redundancy would not be exposed and exploited. Furthermore, the use of the Cholesky factorization allows to maintain the symmetry of $X^T M^{-1} X$.

Linnea uses the following factorizations: Cholesky, LU, QR, symmetric eigenvalue decomposition and singular value decomposition.
$LDL^T$ is currently not supported because with the current LAPACK interface, it is not possible to separately access $L$ and $D$, they can only be used in kernels to directly solve linear systems or invert matrices. Factorizations are only applied to operands that appear inside of the inversion operation, and are not applied to triangular, diagonal and orthogonal operands.

When using matrix factorizations, care has to be taken to avoid infinite loops. As an example, in the expression $S^{-1} B$, it would be possible to first use the Cholesky factorization to factor $S$, resulting in $(L^T L)^{-1} B$, and then compute the product $L^T L$ to obtain the the original expressions. In Linnea, such loops are avoided by labeling operands as factors and by requiring that for any given kernel call, there must be at least one operand that is not a factor.

\subsection{Rewriting Expressions}
\label{sec:rewriting}

In Sec.~\ref{sec:redundancy}, we discussed the conversion of expressions to normal form. In addition, to explore different, algebraically equivalent formulations of a problem, we implemented functions to rewrite expressions into alternative forms. In the algorithm in Fig.~\ref{fig:pseudocode}, this happens in the `representations' function.

Expressions in normal form are rewritten in several ways: We make use of distributivity to convert expressions to the product of sums form. If possible, the inverse operator is pushed up, so $B^{-1} A^{-1}$ is also represented as $(AB)^{-1}$. To explore an even larger set of alternatives, we developed an algorithm to detect common subexpressions of arbitrary length that takes into account identities such as $B^T A^T = (AB)^T$ and $B^{-1} A^{-1} = (AB)^{-1}$. As a result, even terms such as $A^{-1}B$ and $B^T A^{-T}$ are identified as a common subexpression. Since the use of a common subexpression does not necessarily lead to lower computational cost, Linnea also continues to operate on the unmodified expressions. Existing methods for the elimination of redundancy in code, such as common subexpression elimination, partial redundancy elimination, global value numbering \cite[Chap. 13]{Muchnick:1997wv}, are not able to consider algebraic identities.

In addition to those relatively general rewritings, we also encoded a small number of non-trivial rules that allow to compute specific terms at a reduced cost. As an example, $X := A^T A + A^T B + B^T A$ becomes $$Y := B + A/2 \qquad X := A^T Y + Y^T A\text{.}$$ While such transformations are only applicable in special cases, thanks to efficient many-to-one pattern matching, Linnea can identify such cases with only minimal impact on the overall performance.

\subsection{Cost Metric and Pruning}

To select a program that is close to the optimum and satisfies constraints such as memory usage, a cost metric is necessary. This can either be an exact cost or an estimate. Such a metric could consider the number of kernel invocations, the cost of invocations, for example number of floating-point operations, the number of bytes moved or even numerical stability. A suitable cost metric has the additional advantage that it can also be used to prune the search graph: If a branch is known (or expected) not to lead to good solutions, it can be removed from the derivation graph, similar to graph search strategies such as beam search \cite{nilsson2014}.

Presently, as cost metric, Linnea uses the number of floating-point operations (FLOPs). This metric yields estimates for the execution time, and has the advantage that it is easy to determine. For each kernel, Linnea has a formula that computes the number of FLOPs performed by this kernel based on the sizes of the matched operands. As an example, for the \texttt{gemm} kernel, which computes $A B + C$ with $A \in \mathbb{R}^{m \times k}$ and $B \in \mathbb{R}^{k \times n}$, the formula is $2mnk$. Those formulas were either taken from \cite[pp. 336--337]{higham2008}, or inferred by hand. To find the path in the derivation graph with the lowest cost, we use a $K$ shortest paths algorithm \cite{Jimenez:1999cd}.

Linnea is built in a way so that different cost metrics can be used, as long as they are defined over the set of available kernels. However, both performance prediction and automatic stability analysis for linear algebra kernels are known to be challenging problems \cite{Bientinesi:2011bv, Peise:2014fr}. In general, Linnea generates algorithms with different numerical properties.

\subsection{Generation Strategies}
\label{sec:constructive}

For products of multiple matrices, Linnea effectively solves the matrix chain problem \cite[Sec. 15.2]{cormen2009} by enumerating all possible solutions.
For the sum of matrices, all possible parenthesizations are enumerated, even
though it is not expected that different parenthesizations lead to algorithms
that differ significantly in terms of FLOPs. In both cases, this
exhaustive enumeration contributes significantly to the size of the search
graph. To avoid this enumeration, we made use of the flexibility of algebraic code generation and implemented a second, \emph{constructive} generation strategy in Linnea.

This constructive strategy identifies sums and product for which it is possible to generate efficient code by using specialized algorithms.
This change is implemented as an alternative `gen\_suc\-ces\-sors' function:
For products, it uses the generalized matrix chain algorithm \cite{barthels2018cgo}. For sums, we developed a simple greedy algorithm. The disadvantage of those two algorithms is that they can not always make use of the full functionality of kernels, thus potentially leading to suboptimal code.

\subsection{Code Generation}

A path in the derivation graph is only a symbolic representation of an algorithm; it still has to be translated to actual code. Most importantly, all operands are represented symbolically, with no notion of where and how they are stored in memory. During the code generation, operands are assigned to memory locations, and it is decided in which storage format they are stored.

Many BLAS and LAPACK kernels overwrite one of their input operands. As an example, the \texttt{gemm} kernel $\alpha A B + \beta C$ writes the result into the buffer containing $C$. Linnea performs a basic liveness analysis to identify if an operand can be overwritten. If this is not the case, the operand is copied. At present, Linnea does not reorder kernel calls to avoid unnecessary copies.

Some kernels use specialized storage formats for matrices with properties. As an example, for a lower triangular matrix, only the lower, non-zero part is stored. Those storage formats have to be considered when generating code: While specialized kernels for triangular matrices only access the non-zero entries, a more general kernel would read from the entire buffer. Thus, it has to be ensured that operands are always in the correct storage format, if necessary by converting the storage format. Similar to overwriting, storage formats are not considered during the generation of algorithms. During the code generation, operands are converted to different storage formats when necessary. The output of an algorithm is always converted back to the full storage format.

\section{Experiments}
\label{sec:experiments}

To demonstrate Linnea, we perform three different experiments. First, to evaluate the quality of the code generated by Linnea, we compare against
Julia\footnote{Julia 1.1.0-DEV.468 from October 17, 2018.},
Matlab\footnote{Version 2018b.},
Eigen\footnote{Version 3.3.4.},
and Armadillo\footnote{Version 8.500.0.}.
Then, to illustrate the importance of merging
branches, we discuss the generation time with and
without merging branches. Finally, we evaluate the second, constructive generation strategy implemented in Linnea.

The measurements were taken on an Intel Broadwell E5-2650v4 with 2.2 GHz and 128 GB RAM. For all but Matlab, we link against the Intel MKL implementation of BLAS and LAPACK (MKL 2018 update 3) \cite{mkldoc}; Matlab instead uses MKL 2018 update 1.
For the execution of generated code, all reported timings refer to the minimum of 20 repetitions on cold data, to avoid any caching effects. For the generation time, we used one repetition. All code was run single-threaded.

\paragraph{Test Problems}

As tests problems, we use a collection of 25 problems from real applications, including domains such as image and signal processing, statistics, and regularization. A representative selection of those problems is shown in Tab.~\ref{tab:problems}. Operand sizes are selected to reflect realistic use cases.

\subsection{Libraries and Languages}

\begin{table}
\centering
\begin{tabular}{lp{6.1cm}} \toprule
	Name & Implementation \\\midrule
	Julia n & \texttt{inv(A)*B*transpose(C)} \\
	Julia r & \texttt{(A$\backslash$B)*transpose(C)} \\
	Armadillo n & \texttt{arma::inv\_sympd(A)*B*(C).t()} \\
	Armadillo r & \texttt{arma::solve(A, B)*C.t()} \\
	Eigen n & \texttt{A.inverse()*B*C.transpose()} \\
	Eigen r & \texttt{A.llt().solve(B)*C.transpose()} \\
	Matlab n & \texttt{inv(A)*B*transpose(C)} \\
	Matlab r & \texttt{(A$\backslash$B)*transpose(C)} \\\bottomrule
\end{tabular}
\vspace{0.3em}
\caption{Input representations for the expression $A^{-1} B C^T$, where $A$ is SPD and $C$ is lower triangular. The letters `n' and `r' denote the naive and recommended implementation, respectively.
}
\label{tab:implementations}
\end{table}

\newcommand{\LT}{\text{LT}}
\newcommand{\UT}{\text{UT}}
\newcommand{\DI}{\text{DI}}
\newcommand{\SYM}{\text{SYM}}
\newcommand{\SPD}{\text{SPD}}
\newcommand{\SPSD}{\text{SPSD}}
\newcommand{\dims}[3]{$#1 \in \mathbb{R}^{#2 \times #3}$}
\newcommand{\spc}{\;}
\newcommand{\singleline}[3]{{#1} & {#2} & {\small#3}}
\newcommand{\doubleline}[3]{{#1} & \multicolumn{2}{l}{#2} \\ \multicolumn{3}{r}{\small#3}}
\newcommand{\idx}[1]{\parbox{0.4cm}{#1}}

\begin{table*}
\centering
\begin{tabular}{llr}
\toprule

\singleline
{\bf\idx{a)} Generalized Least Squares } 
{\colorbox{rwthverylightgray}{$b := (X^T M^{-1} X)^{-1} X^T M^{-1} y$}}
{\dims{M}{n}{n},\spc\SPD;\spc
\dims{X}{n}{m};\spc
\dims{y}{n}{1};\spc
$n > m$;\spc
$n = 2500$;\spc
$m = 500$}
\\ \midrule

\doubleline
{\bf\idx{b)} Optimization \cite{straszak2015}} 
{\colorbox{rwthverylightgray}{
$x_f := W A^T (AWA^T)^{-1} (b - Ax);\spc
x_o := W (A^T (AWA^T)^{-1} Ax - c)$}}
{\dims{A}{m}{n};\spc
\dims{W}{n}{n},\spc\DI,\spc\SPD;\spc
\dims{b}{m}{1};\spc
\dims{c}{n}{1};\spc
$n > m$;\spc
$n = 2000$;\spc
$m = 1000$}
\\ \midrule

\doubleline
{\bf \idx{c)} Signal Processing \cite{ding2016}} 
{\colorbox{rwthverylightgray}{$x := \left( A^{-T} B^T B A^{-1} + R^T L R \right)^{-1} A^{-T} B^T B A^{-1} y$}}
{\dims{A}{n}{n};\spc
\dims{B}{n}{n};\spc
\dims{R}{n-1}{n},\spc\UT;\spc
\dims{L}{n-1}{n-1},\spc\DI;\spc
\dims{y}{n}{1};\spc
$n = 2000$}
\\ \midrule

\doubleline
{\bf \idx{d)} Triangular Matrix Inv. \cite{bientinesi2008}} 
{\colorbox{rwthverylightgray}{
$X_{10} := L_{10} L_{00}^{-1};\ \ X_{20} := L_{20} + L_{22}^{-1} L_{21} L_{11}^{-1} L_{10};\spc
X_{11} := L_{11}^{-1};\ \ X_{21} := - L_{22}^{-1} L_{21}$}}
{\spc\spc\spc\spc\spc\spc\spc\spc\spc\spc\spc\spc\spc\spc\spc\spc\spc\spc\spc\spc\spc\spc\spc\spc\spc\spc\spc\spc\spc\spc\spc\spc\spc\spc\spc\spc\spc\spc\spc\spc
\dims{L_{00}}{n}{n},\spc\LT;\spc
\dims{L_{11}}{m}{m},\spc\LT;\spc
\dims{L_{22}}{k}{k},\spc\LT;\spc
\dims{L_{10}}{m}{n};\spc
\dims{L_{20}}{k}{n};\spc
\dims{L_{21}}{k}{m};\spc
$n = 2000$;\spc
$m = 200$;\spc
$k = 2000$}
\\ \midrule

\doubleline
{\bf \idx{e)} Ensemble Kalman Filter \cite{nino2016}} 
{\colorbox{rwthverylightgray}{$X^a := X^b + \left( B^{-1} + H^T R^{-1} H \right)^{-1} \left( Y - H X^b \right)$}}
{\dims{B}{N}{N} \SPSD;\spc
\dims{H}{m}{N};\spc
\dims{R}{m}{m} \SPSD;\spc
\dims{Y}{m}{N};\spc
\dims{X^b}{n}{N};\spc
$N = 200$;\spc
$n = 2000$;\spc
$m = 2000$}
\\ \midrule

\doubleline
{\bf \idx{f)} Image Restoration \cite{Tirer:2017uv}} 
{\colorbox{rwthverylightgray}{$x_k := (H^T H + \lambda \sigma^{2} I_{n} )^{-1} ( H^T y +\lambda \sigma^{2}(v_{k-1} - u_{k-1}) )$}}
{\dims{H}{m}{n};\spc
\dims{y}{m}{1};\spc
\dims{v_{k-1}}{n}{1};\spc
\dims{u_{k-1}}{n}{1};\spc
$\lambda > 0$;\spc
$\sigma > 0$;\spc
$n > m$;\spc
$n = 5000$;\spc
$m = 1000$}
\\ \midrule

\doubleline
{\bf \idx{g)} Rand. Matrix Inversion \cite{Gower:2017bq}} 
{\colorbox{rwthverylightgray}{
$\Lambda := S (S^T A^T W A S)^{-1} S^T;\spc X_{k+1} := X_k + (I_n - X_k A^T) \Lambda A^T W$}}
{\dims{W}{n}{n},\spc\SPD;\spc
\dims{S}{n}{q};\spc
\dims{A}{n}{n};\spc
\dims{X_k}{n}{n};\spc
$q \ll n$;\spc
$n = 5000$;\spc
$q = 500$}
\\  \midrule

\doubleline
{\bf \idx{h)} Rand. Matrix Inversion \cite{Gower:2017bq}} 
{\colorbox{rwthverylightgray}{$ X_{k+1} := S(S^T A S)^{-1} S^T + (I_n - S(S^T A S)^{-1} S^T A) X_k (I_n - A S(S^T A S)^{-1} S^T)$}}
{\dims{A}{n}{n},\spc\SPD;\spc
\dims{W}{n}{n},\spc\SPD;\spc
\dims{S}{n}{q};\spc
\dims{X_k}{n}{n};\spc
$q \ll n$;\spc
$n = 5000$;\spc
$q = 500$}
\\ \midrule

\doubleline
{\bf \idx{i)} Stoch. Newton \cite{Chung:2017ws}} 
{\colorbox{rwthverylightgray}{$ B_k := \frac{k}{k-1}B_{k-1} (I_n - A^T W_k ((k-1)I_l + W_k^T A B_{k-1} A^T W_k)^{-1} W_k^T A B_{k-1} )$}}
{\dims{W_k}{m}{l};\spc
\dims{A}{m}{n};\spc
\dims{B_k}{n}{n},\spc\SPD;\spc
$l < n \ll m$;\spc
$l = 625$;\spc
$n = 1000$;\spc
$m = 5000$}
\\ \midrule

\singleline
{\bf \idx{j)} Tikhonov reg. \cite{Golub:2006hl}} 
{\colorbox{rwthverylightgray}{$x := (A^T A + \Gamma^T \Gamma)^{-1} A^T b$}}
{\dims{A}{n}{m};\spc
\dims{\Gamma}{m}{m};\spc
\dims{b}{n}{1};\spc
$n = 5000$;\spc
$m = 50$}
\\ \midrule

\doubleline
{\bf \idx{k)} Gen. Tikhonov reg.  } 
{\colorbox{rwthverylightgray}{$x := (A^T P A + Q)^{-1} (A^T P b + Q x_0 )$}}
{\dims{P}{n}{n},\spc\SPSD;\spc
\dims{Q}{m}{m},\spc\SPSD;\spc
\dims{x_0}{m}{1};\spc
\dims{A}{n}{m};\spc
\dims{\Gamma}{m}{m};\spc
\dims{b}{n}{1};\spc
$n = 5000$;\spc
$m = 50$}
\\ \midrule

\doubleline
{\bf \idx{l)} LMMSE estimator \cite{Kabal:2011wr}} 
{\colorbox{rwthverylightgray}{$x_\text{out} = C_X A^T (A C_X A^T + C_Z)^{-1} (y - A x) + x$}}
{\dims{A}{m}{n};\spc
\dims{C_X}{n}{n},\spc\SPSD;\spc
\dims{C_Z}{m}{m},\spc\SPSD;\spc
\dims{x}{n}{1};\spc
\dims{y}{m}{1};\spc
$n = 2000$;\spc
$m = 1500$}
\\ \midrule

\doubleline
{\bf\idx{m)} Kalman Filter \cite{Kalman:1960ii}} 
{\colorbox{rwthverylightgray}{
$K_k  := P_k^b H^T ( H P_k^b H^T + R )^{-1};\spc
x_k^a := x_k^b + K_k ( z_k - H x_k^b );\spc
P_k^a := \left( I - K_K H \right) P^b_k$}}
{\dims{K_k}{n}{m};\spc
\dims{P_k^b}{n}{n},\spc\SPD;\spc
\dims{H}{m}{n},\spc\SPD;\spc
\dims{R}{m}{m},\spc\SPSD;\spc
\dims{x_k^b}{n}{1};\spc
\dims{z_k}{m}{1};\spc
$n = 400$;\spc
$m = 500$}
\\ \bottomrule

\end{tabular}
\caption{A selection of the 25 application problems used in the experiments. Matrix properties: diagonal (DI), lower/upper triangular (LT/UT), symmetric positive definite (SPD), symmetric positive semi-definite (SPSD), symmetric (SYM).}
\label{tab:problems}
\end{table*}

For each library and language, two different implementations are used:
\emph{naive} and \emph{recommended}. The naive implementation is the one that
comes closest to the mathematical description of the problem. It is also closer to
input to Linnea.
As an example, in Julia, the naive implementation for $A^{-1} B$ is \texttt{inv(A)*B}. However, since documentations almost always discourage this use of the inverse operator, we also consider a so called \emph{recommended} implementation, which uses dedicated functions to solve linear systems (\texttt{A$\backslash$B}).

In the following, we describe the different implementations. As examples, in Tab.~\ref{tab:implementations} we provide the implementations of $A^{-1} B C^T$, where $A$ is symmetric positive definite and $C$ is lower triangular.

\begin{description}
	\item[Julia] Properties are expressed via types. The naive
          implementation uses \texttt{inv()}, while the recommended one uses
          the $\slash$ and $\backslash$ operators.
	\item[Matlab] The naive implementation uses \texttt{inv()}, the
          rec\-om\-mend\-ed one the $\slash$ and $\backslash$ operators.
	\item[Eigen] The recommended implementation selects linear systems
          solvers based on the matrix properties, and uses views to describe properties.
	\item[Armadillo] In the naive implementation, specialized
          functions are used for the inversion of SPD and diagonal
          matrices.
          For \texttt{solve}, we use the \texttt{solve\_opts::fast}
          option to disable an expensive refinement. In addition,
          \texttt{trimatu} and \texttt{trimatl} are used for triangular matrices.
\end{description}

\subsection{Results}

\newcommand{\markscaling}{1}
\newcommand{\markscalingsmall}{0.7}

\pgfplotstableread[col sep=comma]{speedup_ce.csv}\applicationspeedup

\pgfplotstablesort[sort key=mean, sort cmp=float <]\applicationspeedups{\applicationspeedup}

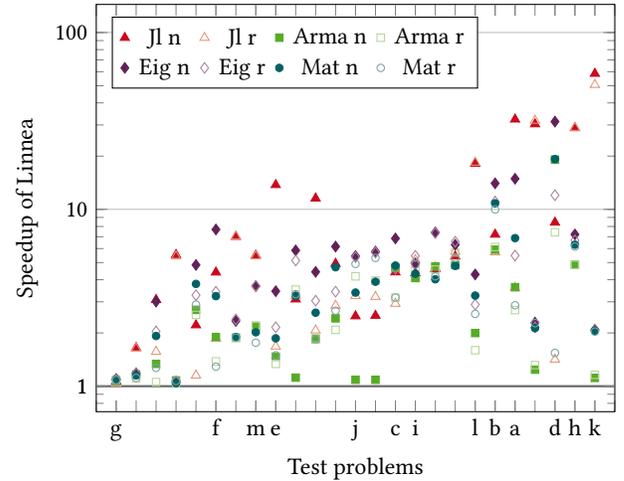
\begin{figure}[]
    \centering
    \begin{tikzpicture}
        \begin{axis}[
        	extra x ticks={20, 19, 14, 22, 8, 5, 0, 23, 15, 12, 24, 18, 7},
        	extra x tick labels={ a, b, c, d, e, f, g, h, i, j, k, l, m},
        	extra x tick style={
        		draw=none,
        		grid=none,
        		font=\normalsize,
        		text height=1.5ex,
        		anchor=north
        	},
            height=7cm,
            width=\columnwidth,
            xlabel={Test problems},
            ylabel={Speedup of Linnea},
            grid=major,
            ymode=log,
            ymax=100,
			xmin=-1,
			xmax=25,
            enlarge y limits=0.08,
            xtick distance=1,
			xmajorgrids=false,
			xticklabel=\empty,
			extra y ticks={1},
			extra y tick style={
				grid = major,
				grid style={
					color=rwthblack75,
					thick,
				}
			},
            legend style={
                column sep=0.2em,
            },
			legend pos=north west,
            legend columns=4,
            log ticks with fixed point,
        ]
            \addplot[
                fill=rwthred,
                draw=rwthred,
                mark=triangle*,
                only marks,
                mark options={scale=\markscaling}
            ]
                table[
                    x expr=\coordindex,
                    y=naive_julia,
                ] {\applicationspeedups};
            \addlegendentry{Jl n};
            \addplot[
                fill=rwthred50,
                draw=rwthred50,
                mark=triangle,
                only marks,
                mark options={scale=\markscaling}
            ]
                table[
                    x expr=\coordindex,
                    y=recommended_julia,
                ] {\applicationspeedups};
            \addlegendentry{Jl r};
            \addplot[
                fill=rwthgreen,
                draw=rwthgreen,
                mark=square*,
                only marks,
                mark options={scale=\markscalingsmall}
            ]
                table[
                    x expr=\coordindex,
                    y=naive_armadillo,
                ] {\applicationspeedups};
            \addlegendentry{Arma n};
            \addplot[
                fill=rwthgreen50,
                draw=rwthgreen50,
                mark=square,
                only marks,
                mark options={scale=\markscalingsmall}
            ]
                table[
                    x expr=\coordindex,
                    y=recommended_armadillo,
                ] {\applicationspeedups};
            \addlegendentry{Arma r};
            \addplot[
                fill=rwthviolet,
                draw=rwthviolet,
                mark=diamond*,
                only marks,
                mark options={scale=\markscaling}
            ]
                table[
                    x expr=\coordindex,
                    y=naive_eigen,
                ] {\applicationspeedups};
            \addlegendentry{Eig n};
            \addplot[
                fill=rwthviolet50,
                draw=rwthviolet50,
                mark=diamond,
                only marks,
                mark options={scale=\markscaling}
            ]
                table[
                    x expr=\coordindex,
                    y=recommended_eigen,
                ] {\applicationspeedups};
            \addlegendentry{Eig r};
            \addplot[
                fill=rwthpetrol,
                draw=rwthpetrol,
                mark=*,
                only marks,
                mark options={scale=\markscalingsmall}
            ]
                table[
                    x expr=\coordindex,
                    y=naive_matlab,
                ] {\applicationspeedups};
            \addlegendentry{Mat n};
            \addplot[
                fill=rwthpetrol50,
                draw=rwthpetrol50,
                mark=o,
                only marks,
                mark options={scale=\markscalingsmall}
            ]
                table[
                    x expr=\coordindex,
                    y=recommended_matlab,
                ] {\applicationspeedups};
            \addlegendentry{Mat r};
        \end{axis}
    \end{tikzpicture}
    \caption{Speedup of Linnea over other languages and libraries for 25 application problems. The problems are sorted by the average of all speedups. The labels on the x-axis correspond to the labels of the 13 problems shown in Tab.~\ref{tab:problems}.}
    \label{fig:speedupapplication}
\end{figure}

For 22 out of 25 test problems, Linnea's generation time lies between $0.2$ seconds and $1$ hour. In 9 cases, the generation time is below $1$ minute, in 17 cases, it is below $10$ minutes. For three of the application problems
(including h) and i) in Tab.~\ref{tab:problems}), Linnea's default exhaustive strategy does not find a solution within 2 hours, so the constructive strategy is used instead.

In Fig.~\ref{fig:speedupapplication}, we present the speedup of the code generated by Linnea over other languages and libraries for the application test cases. For all test problems Linnea generates the fastest algorithm.
To understand where the speedups for the code generated by Linnea come
from, we discuss the details of few exemplary test problems.

\paragraph{Distributivity}

The assignments
\begin{align*}
	H^\dag &:= H^T ( H H^T )^{-1} \text{,} \\
	y_k &:= H^\dag y + ( I_n - H^\dag H ) x_k\text{,}
\end{align*}
which are part of an image restoration application \cite{Tirer:2017uv}, illustrate well
how distributivity might affect performance.
Due to the matrix-matrix product $H^\dag H$, computing $y_k$ based on the original formulation of the problem requires $\mathcal{O}(n^3)$ FLOPs. Instead, Linnea finds the solution
\begin{align*}
	v_\text{tmp} &:= - H x_k + y \\
	y_k &:= H^\dag v_\text{tmp} + x_k,
\end{align*}
which only uses matrix-vector products (\texttt{gemv}), and requires $\mathcal{O}(n^2)$ FLOPs.
This solution is obtained in two steps: First, $H^\dag y + ( I_n - H^\dag H ) x_k$ is converted to Linnea's normal form, returning $H^\dag y + x_k - H^\dag H x_k$; then, by factoring out $H^\dag$, the expression is written back as product of sums, resulting in $H^\dag (y - H x_k) + x_k$, which can be computed with two calls to \texttt{gemv}.
Here, this optimization yields speedups between 4\su{} (Matlab naive) and 7.4\su{} (Eigen naive) with respect to the other languages and libraries.

\paragraph{Associativity} With the exception of Armadillo, none of the languages and libraries we compare with consider the matrix chain problem. Instead, products are always computed from left to right. The expression $x := W(A^T(AWA^T)^{-1}b-c)$, which comes from an optimization problem \cite{straszak2015}, is a good example to illustrate the importance of making use of associativity in products. We assume that the subexpression $S = AWA^T$ was already computed. Since $S$ is a full matrix, a factorization has to be applied to solve the linear; since $S$ is symmetric positive definite, it is possible to apply the Cholesky factorization.  Thus, the expression $W(A^T L^{-T} L^{-1} b-c)$ is left to compute. Regardless of whether this expression is represented as $W(A^T L^{-T} L^{-1} b-c)$ or $W A^T L^{-T} L^{-1} b - W c$, a left-to-right evaluation requires the solution of two lower triangular linear systems with multiple right-hand sides, which have cubic complexity.  By contrast, the solution found by Linnea, which is shown in Fig.~\ref{fig:codeassociativity}, only uses operations with quadratic complexity: two lower triangular linear systems with a single right-hand side (\texttt{trsv}), a matrix-vector product (\texttt{gemv}), and a product of a diagonal matrix with a vector, which is implemented as an elementwise product of the diagonal elements and the vector. In addition, this sequence of four kernels does not require any additional memory: variables \texttt{b} and \texttt{c} both get overwritten. The speedup for the entire expression is between 1.15\su{} and 4.8\su{}.

\begin{figure}
\lstset{
	basicstyle=\ttfamily,
	numbers=left,
	xleftmargin=2em,
	emph={end},
	emphstyle=\bfseries,
}
\hrule
\begin{lstlisting}
W = diag(W)
Acopy = Array{Float64}(undef, 1000, 2000)
blascopy!(1000*2000, A, 1, Acopy, 1)
for i = 1:size(A, 2);
    view(A, :, i)[:] .*= W[i];
end;
S = Array{Float64}(undef, 1000, 1000)
gemm!('N', 'T', 1.0, A, Acopy, 0.0, S)
potrf!('L', S)
trsv!('L', 'N', 'N', S, b)
trsv!('L', 'T', 'N', S, b)
gemv!('T', 1.0, Acopy, b, -1.0, c)
c .*= W
\end{lstlisting}
\hrule
\caption{The generated code for $x := W(A^T(AWA^T)^{-1}b-c)$. Variables were renamed for better readability. Lines 4--6 is one of the code snippets for operations not supported by BLAS and LAPACK; the multiplication of a full and a diagonal matrix.}
\label{fig:codeassociativity}
\end{figure}

\paragraph{Common Subexpressions} Expressions arising in application
frequently exhibit common subexpressions; one such example is given by the assignment $$B_1 := \frac{1}{\lambda_1} (I_n - A^T W_1 (\lambda_1 I_l + W_1^T A A^T W_1)^{-1} W_1^T A )\text{,}$$ which is used in the solution of large least-squares problems \cite{Chung:2017ws}. 
Linnea successfully identifies that the term $W_1^T A$, or in transposed form
$(A^T W_1)^T$, appears four times and computes this subexpression only
once. In this example, these savings lead to speedups between 4.7\su{} and 6.6\su{}.

\paragraph{Properties} Many matrix operations can be sped up by taking advantage of matrix properties. As an example, here we discuss the evaluation of the assignment $x := (A^T A + \alpha^2 I)^{-1} A^T b$, a least-squares problem with Tikhonov regularization \cite{Golub:2006hl}, where matrix $A$ is of size $5000 \times 50$ and has full rank. Since $A$ has more rows than columns and is full rank, Linnea is able to infer that $A^T A$ is not only symmetric, but also positive definite (SPD). Similarly, Linnea infers that $\alpha^2 I$ is SPD because 1) the identity matrix is SPD, 2) $\alpha^2$ is positive and 3) a SPD matrix scaled by a positive factor is still SPD. Since the sum of two SPD matrices is SPD, $A^T A + \alpha^2 I$ is identified as SPD. As a result, the Cholesky factorization is used to solve the linear system. If $A^T A + \alpha^2 I$ had not been identified as SPD, a more expensive factorization such as LU had to be used. Finally, since Linnea infers properties based on the annotations of the input matrices, no property checks have to be performed at runtime; if the input matrices have the specified properties, all inferred properties hold. Altogether, the code generated for this assignment is between 1.08\su{} and 5.8\su{} faster than the other languages and libraries.

In general, the speedup of Linnea depends both on the potential for optimization in a given problem, as well as on the similarity of the default evaluation strategy in each language and library to the optimal one.

In case of problem j) for example, the naive Armadillo implementation is only 1.09\su{} slower than the code generation by Linnea, while all other implementations are around 5\su{} slower. The reason is that for this problem, the parenthesization has the largest influence on the execution time. While Armadillo does solve a simplified version of the matrix chain problem, the \texttt{solve} function used in the recommended implementation (see Tab.~\ref{tab:implementations}) effectively introduces a fixed parenthesization.
Due to the explicit inversion in the naive implementation, there is no such fixed parenthesization, so Armadillo is able to find a solution which is faster, but numerically less stable, than the solution generated by Linnea.

For problem d), which is the loop body of a blocked algorithm for the inversion of a triangular matrix, there is a large spread between the speedups: The recommended Julia and Matlab solutions are respectively around 1.40\su{} and 1.50\su{} slower than Linnea, while the naive Armadillo and Eigen implementations are 20\su{} and 30\su{} slower.
Instead, this spread is likely caused by a combination of the interface the different systems offer, and how they utilize properties.
Both Armadillo and Eigen do not have functions to solve linear systems of the form $A B^{-1}$, with the inverted matrix on the right-hand side. Thus, even in the recommended solution, for $X_{10} := L_{10} L_{00}^{-1}$, explicit inversion is used instead. Most likely, Armadillo and Eigen are not able to identify that $L_{00}$ is lower triangular and instead use an algorithm for the inversion of a general matrix, leading to a significant loss in performance, while Julia and Matlab correctly use the \texttt{trsm} kernel.

For expression g), all solutions have very similar execution times; the speedup of Linnea is between 3\% and 9\%. The cost of computing this problem is dominated by the cost of computing the value of $X_{k+1}$, for which the solution found by all other languages and libraries is almost identical to the solution found by Linnea. While Linnea is able to save some FLOPs in the computation of $\Lambda$, those savings are negligible for the evaluation of the entire problem.

\subsubsection{Impact of Merging Branches}
As discussed in Sec.~\ref{sec:redundancy}, in order to reduce the size of the search graph and thus to speed up the program generation, without affecting the quality of the solutions, different branches of the derivation graph are merged.
In our experiments, we observe that merging reduces the number of nodes, and likewise, the generation time, by up to two orders of magnitude. Typically, the reduction of the derivation time is larger for larger graphs. We observed that only for very simple input problems with a generation time of less than 0.1 seconds, the overhead due to merging branches slows down the derivation.

\subsubsection{Constructive Strategy}

\pgfplotstableread[col sep=comma]{generation_time.csv}\randomtime
\pgfplotstablesave[string replace={NaN}{7200}]{\randomtime}{tmp.txt}
\pgfplotstableread[]{tmp.txt}\randomtimeR
\pgfplotstablesort[sort key=constructive_merging,sort cmp=float <]\randomtimesorted{\randomtimeR}

\newcommand{\markscalingtime}{.8}

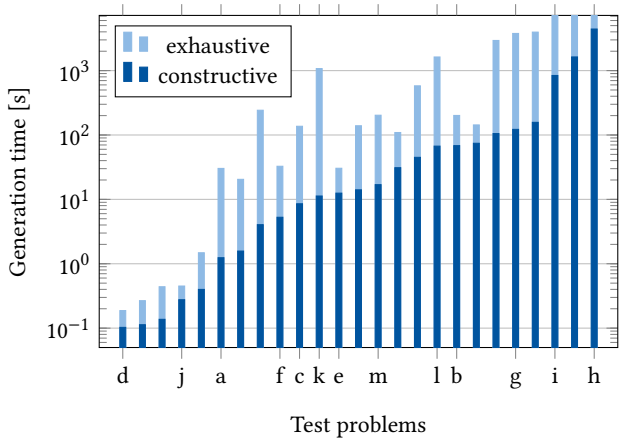
\begin{figure}[]
    \centering
    \begin{tikzpicture}[]
        \begin{axis}[
            height=6cm,
            width=\columnwidth,
            xlabel={Test problems},
            ylabel={Generation time [s]},
            grid=major,
            ybar=0pt,
            bar width=0.3,
            bar shift=0,
            xtick=\empty,
            ymode=log,
            ymin=0.05,
            ymax=7200,
            log origin=infty,
            enlarge x limits=0.05,
            extra x ticks={5, 17, 9, 0, 11, 8, 20, 24, 22, 3, 10, 16, 13},
        	extra x tick labels={ a, b, c, d, e, f, g, h, i, j, k, l, m},
        	extra x tick style={
        		draw=none,
        		grid=none,
        		font=\normalsize,
        		text height=1.5ex,
        		anchor=north
        	},
            legend style={
            	column sep=0.2em,
            },
			legend pos=north west,
            legend columns=1,
        ]
            \addplot[
            	fill=rwthblue50,
            	draw=rwthblue50,
            	mark options={scale=\markscalingtime}
            ]
                table[
                    x expr=\coordindex,
                    y=exhaustive_merging,
                ] {\randomtimesorted};
            \addlegendentry{exhaustive};
            \addplot[
            	fill=rwthblue,
            	draw=rwthblue,
            	mark options={scale=\markscalingtime}
            ]
                table[
                    x expr=\coordindex,
                    y=constructive_merging,
                ] {\randomtimesorted};
            \addlegendentry{constructive};
        \end{axis}
    \end{tikzpicture}
    \caption{Generation time for both strategies (sorted by the time for the constructive
      strategy). For the three rightmost problems, the exhaustive strategy does not find a
      solution within two hours. The labels on the x-axis correspond to the labels in Tab.~\ref{tab:problems}.}
    \label{fig:constructivetime}
\end{figure}

To evaluate the effectiveness of the constructive strategy, we used it for the same test set as before. The generation time for both strategies is shown in Fig.~\ref{fig:constructivetime}. Compared to the exhaustive  strategy, the constructive one reduces the generation time by up to two orders of magnitude, and finds a solution for all test problems in at most 70 minutes. In terms of execution time, the code generated by the constructive strategy is on average about 10\% slower.

As an example for the difference between the code generated by the two strategies, consider the assignment
$$x_k := (H^T H + \lambda \sigma^{2} I_{n} )^{-1} ( H^T y +\lambda \sigma^{2}(v_{k-1} - u_{k-1}) )\text{,}$$ which appears in an image restoration application (example 10).
We use $\tau$ for the result of $\lambda \sigma^{2}$. With the exhaustive strategy, the subexpression $H^T H + \tau I_{n}$ is computed by a single call to the BLAS kernel \texttt{syrk}.
With the constructive strategy, the same subexpression is computed in two steps, with one call to \texttt{syrk} and one to \texttt{axpy}:
\begin{align*}
	M_\text{tmp1} &:= H^T H \\
	M_\text{tmp2} &:= M_\text{tmp1} + \tau I_n
\end{align*}
Overall, the algorithm generated by the default strategy uses 9 kernel invocations, while the one generated by the constructive strategy uses 12. While the code generated by the constructive strategy is about 1.7\% slower, the generation time is reduced by a factor of 6, from 33 to 5 seconds.

\section{Conclusion and Future Work}
\label{sec:conclusion}

We presented Linnea, a code generator that translates a high-level linear algebra problem into an efficient sequence of high-per\-for\-mance kernels. In contrast to other languages and libraries, Linnea uses domain knowledge such as associativity, commutativity, distributivity and matrix properties to derive efficient algorithms.
Our experiments on application problems indicate that Linnea outperforms all the current state-of-the-art tools. Linnea is also very flexible, allowing us to either exhaustively
explore the search space, or to quickly find good, but potentially not optimal solutions.

In the future, we aim to integrate the expected efficiency and scalability of kernels into the cost function, and  investigate the generation of parallel code, the extension to sparse linear algebra, as well as ways to speed up the code generation.

\begin{acks}                            
Financial support from the Deutsche Forschungsgemeinschaft (German Research
Foundation) through grants GSC 111 and IRTG 2379 is gratefully
acknowledged. We thank Jan Vitek and Marcin Copik for their help.
\end{acks}

%

%
\bibliographystyle{ACM-Reference-Format}
\bibliography{PhD_bibliography,PhD_bibliography_papers,my_publications}

\end{document}